\newcommand{\algaas}{$\rm Al_{x}\-Ga_{1-x}\-As$}
\newcommand{\gaasalg}{$\rm GaAs/Al_{x}\-Ga_{1-x}\-As$}
\newcommand{\voc}{$\rm V_{oc}$}
\newcommand{\jsc}{$\rm J_{sc}$}
\newcommand{\pin}{$p~-~i~-~n\ $}
\documentstyle[epsf]{elsart}
\begin{document}

\parskip 0.15ex 

\begin{frontmatter}
                
\title{Optimisation of High Efficiency
 $\rm Al_{x}Ga_{1-x}As$ MQW Solar Cells}
\author{J.P.Connolly, K.W.J.Barnham, J.Nelson, P. Griffin, G. Haarpaintner}
\address{Blackett Laboratory, Imperial College of Science,
         Technology and Medicine, London SW7 2BZ}       
\author{C.Roberts}
\address{IRC for Semiconductor Materials, Imperial College of Science,
         Technology and Medicine, London SW7 2BZ}
\author{M.Pate, J.S.Roberts}
\address{EPSRC III-V Facility, University of Sheffield,
         Sheffield S1 3JD}

\begin{abstract}

The \gaasalg\ materials 
system is well suited to multi-bandgap applications
such as the multiple quantum well  solar cell. GaAs
quantum wells are inserted in the undoped \algaas\
active region of a \pin\ structure to extend the
absorption range while retaining a higher open circuit
voltage than would be provided by a cell made of the
well material alone. Unfortunately aluminium
gallium arsenide ($\rm Al_{x}\-Ga_{1-x}\-As$)
suffers from poor transport characteristics
due to DX centres and oxygen contamination during growth,
which degrade the spectral response.
We investigate three mechanisms for improving the spectral response
of the MQW solar cell while an experimental study
of the open circuit voltage examines the voltage enhancement.
An optimised structure for a high efficiency \gaasalg\ solar cell
is proposed.

\end{abstract}

\end{frontmatter}

\section{Introduction}

\begin{figure*}
\label{celldesign}
\vspace{-3ex}
\setlength{\unitlength}{0.88mm}
\begin{picture}(100,120)(-40,-8)

\put(8,87){\makebox(0,0){\small window}}
\put(30,76){\makebox(0,0){\small graded \normalsize $p$}}
\put(65,57){\makebox(0,0){$i$}}
\put(97,40){\makebox(0,0){$n$}}
\put(19,-2){\vector(-1,0){11}}
\put(20,-3){\makebox(0,0)[bl]{$x_{p}$}}
\put(24,-2){\vector(1,0){9}}
\put(38,-3){\makebox(0,0)[bl]{$x_{wp}$}}
\put(58,-2){\vector(-1,0){13}}
\put(60,-3){\makebox(0,0)[bl]{$x_{i}$}}
\put(65,-2){\vector(1,0){12}}
\put(80,-3){\makebox(0,0)[bl]{$x_{wn}$}}
\put(94,-2){\vector(-1,0){5}}
\put(96,-3){\makebox(0,0)[bl]{$x_{n}$}}
\put(100,-2){\vector(1,0){4}}

\put(0,0){\vector(1,0){115}}
\put(115,-5){\makebox(0,0){Depth}}
\put(0,0){\vector(0,1){90}}
\put(-9.5,85){\makebox(0,0){electron}}
\put(-9,80){\makebox(0,0){energy}}

\put(8,-1){\line(0,1){2}}
\put(35,-1){\line(0,1){2}}
\put(45,-1){\line(0,1){2}}
\put(77,-1){\line(0,1){2}}
\put(87,-1){\line(0,1){2}}
\put(107,-1){\line(0,1){2}}


\put(9.2,82.9){\line(-1,0){8}}
\put(9.2,74.9){\line(0,1){8}}
\put(24.3,72.4){\line(-6,1){15}}
\put(39,67.4){\line(-3,1){15}}
\put(39,67.4){\line(5,-2){4}}
\put(53,59.3){\line(-3,2){10}}
\put(53,59.3){\line(0,-1){6}}
\put(57,50.5){\line(-3,2){4}}
\put(57,50.5){\line(0,1){6}}
\put(61,54){\line(-3,2){4}}
\put(61,54){\line(0,-1){6}}
\put(65,45.4){\line(-3,2){4}}
\put(65,45.3){\line(0,1){6}}
\put(69,48.7){\line(-3,2){4}}
\put(69,48.7){\line(0,-1){6}}
\put(73,40.1){\line(-3,2){4}}
\put(73,40.1){\line(0,1){6}}
\put(83,39.4){\line(-3,2){10}}
\put(82.98,39.4){\line(3,-1){4}}
\put(87,38){\line(1,0){20}}


\put(39,31.4){\line(-1,0){37}}
\put(39,31.4){\line(3,-1){4}}
\put(53,23.3){\line(-3,2){10}}
\put(53,29.3){\line(0,-1){6}}
\put(57,26.5){\line(-3,2){4}}
\put(57,20.5){\line(0,1){6}}
\put(61,18){\line(-3,2){4}}
\put(61,24){\line(0,-1){6}}
\put(65,21.4){\line(-3,2){4}}
\put(65,15.3){\line(0,1){6}}
\put(69,12.7){\line(-3,2){4}}
\put(69,18.7){\line(0,-1){6}}
\put(73,16.1){\line(-3,2){4}}
\put(73,10.1){\line(0,1){6}}
\put(83,3.4){\line(-3,2){10}}
\put(82.98,3.4){\line(3,-1){4}}
\put(87,2){\line(1,0){20}}

\multiput(3,34)(10,0){11}{\line(1,0){5}}
\put(113,34){\makebox(0,0){$E_{F}$}}

\end{picture}
\caption{QWSC band diagramme. The structure
is a \protect \pin\ photodiode
design with quantum wells in the intrinsic region. A high band-gap
\protect \algaas\ window is grown on the top surface to reduce
surface recombination.}
\end{figure*}
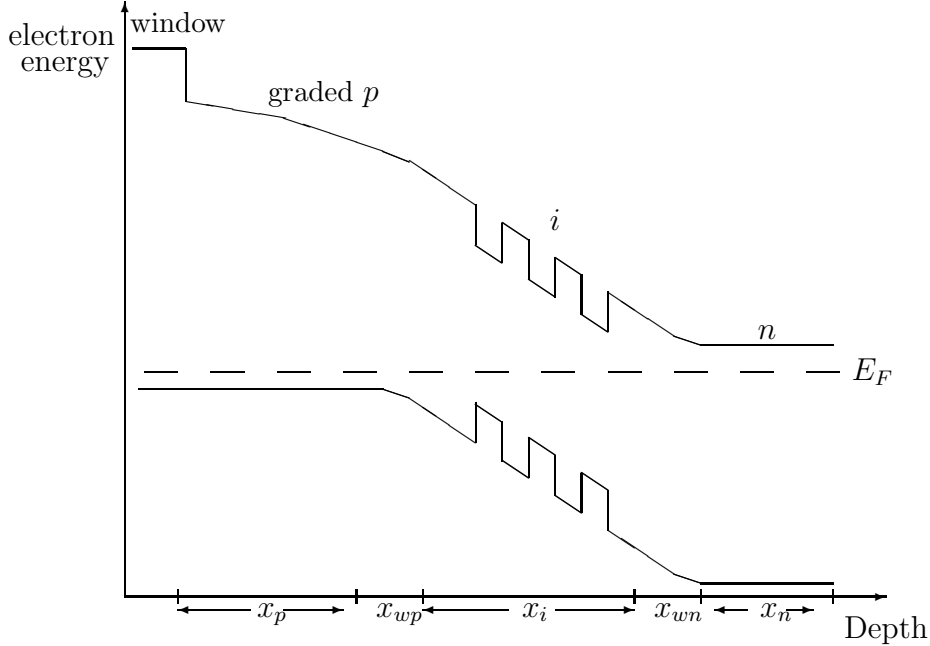

One of the best materials for single band-gap solar cells
is GaAs. Single crystal GaAs has high minority carrier
mobility and a direct bandgap close to the optimum for
single band-gap solar cells. A novel high efficiency design,
the quantum well solar cell (QWSC),
is illustrated in figure \ref{celldesign}.
The multiple quantum wells (MQWs) in the depletion region
extend the absorption range below the barrier band-gap. 
Comparison of $\rm AL_{0.3}Ga_{0.7}As$  devices with and without
quantum wells has shown that the \jsc\ is more than doubled
in samples with 50 MQWs. Furthermore,
the \voc\ provided by QWSCs is higher than would
expected from a cell with the quantum well effective
bandgap. This provides the potential for a QWSC cell with a
higher fundamental efficiency limit than a single band-gap
cell. In order to test this idea in the \gaasalg\ materials
system the short circuit current ($J_{sc}$) must be increased
further. This can be achieved through improved growth and
cell design.

Bulk \algaas\ cells generally suffer from a relatively poor photon
to photocurrent conversion effeciency (quantum efficiency or QE)
Part of this poor performance stems from
a DX centre associated with the
proximity in energy of the $\Gamma$, $X$ and $L$ band gaps in
energy at approximately 35\% aluminium (Al) \cite{singh93}.
Oxygen contamination from the Al sources during
growth contributes further to decreasing minority carrier 
lifetimes with increasing Al fraction $X$.

Three methods of optimising the QE
are described. The first is investigated in \cite{paxman93} 
and consists of thinning the $p$ layer thickness. 
This improves minority carrier collection
by reducing the mean distance carriers must diffuse before
reaching the junction. Moreover, the absorptivity of
the $p$ layer is reduced, allowing more light to reach the
efficient intrinsic ($i$) region. However, the
\jsc\ current enhancement is offset by
increased $p$ layer series resistance.

The second method consists of linearly grading $X$
in the $p$ layer so that the bandgap decreases with depth. This
also reduces the $p$ layer absorptivity, and enhances the $i$
region photocurrent. Furthermore, minority carriers
generated in the $p$ are swept towards the junction with the intrinsic
region by the band-gap gradient. This only affects the QE at energies
above the \algaas\ band-gap. At very high Al content, carrier
collection efficiency tends to decrease because of a deterioration
in material quality.

The third improvement consists of etching off the
GaAs substrate and coating the back of the cell with a
mirror. This is very effective for a QWSC design because
the wells absorb a relatively small fraction of the incident
light but convert it into current with nearly
100\% efficiency.

The following discussion outlines the voltage study described
in \cite{haarpaintner94}. Thinner $p$ regions are described
in \cite{paxman93} while the present work mainly investigates samples
with graded layers and mirrors. Experimental and modelling results
incorporating these methods are presented, with particular
regard to to QE enhancement in samples with graded $p$ layers
and mirror backed samples.
The model is used to design an optimised structure for a 50
quantum well QWSC.
     
\section{Voltage}
\label{voltage}

Measurements in \cite{haarpaintner94} 
have shown that the QWSC \voc\ as a function of effective band-gap
is higher than expected from detailed balance arguments.
This has been seen in the \gaasalg\, indium gallium arsenide / indium
phosphide and also the indium gallium phosphide / gallium
arsenide materials systems. Photoluminescence and photocurrent
studies of \algaas\ samples presented in \cite{QFL}
indicate that the quasi-Fermi level
separation in the wells is also greater than expected. This
phenomenon is not fully understood but may be due to the high
thermal escape efficiencies observed at room temperature.

The observed voltage in a ungraded QWSC with 20\% Al
fraction in the barrier $X_{barr}$ was 1\% higher than exptected
for a cell with the
well bandgap. Similar voltage enhancements in cells with
higher Al fraction were 7\% with a $X_{barr}$=30\% and 11\%
with $X_{barr}$=40\%. However, \voc\ optimisation is limited
by the decreasing QE at high $X_{barr}$.  

\section{Theory}
\label{theory}

\subsection{Quantum Efficiency}
\label{QEsection}

The QE model differs from previous work by
\cite{paxman93} and \cite{hovel75} by considering inhomogeneous
material with position dependent materials parameters. We compare
the modelled QE with experimental data for graded $p$ layer devices.
The QE of the cell is calculated by solving
minority carrier transport equations at room temperature.
The calculation of photocurrent from doped layers
applies equally to $p$ and $n$ regions, with appropriate
materials parameters. The $n$ region contribution is
small because little light with sufficient energy
reaches it. For this region, the following discussion
concentrates on photocurrent contributions from the $p$
and $i$ layers.
Under illuminated conditions, the excess minority carrier generation
rate as a function of depth $x$ from the surface of the $p$ layer
is given by

\begin{equation}  
\label{GenerationRate}
\begin{array}{l l}                         
G(x,\lambda) =
& F(\lambda)(1-R(\lambda))\alpha(x,\lambda)             \\
& \times exp (-\int_{0}^{x} \left[ \alpha(x,\lambda) \right] dx) \\
\end{array}
\end{equation}

where $R(\lambda)$ is the surface reflectivity, 
$\alpha(x,\lambda)$ is the absorption coefficient
and $F(\lambda)$ is the incident photon flux.
Current and continuity equations determine the excess
minority carrier concentration $n(x)$. Since the cell
operates in the low injection limit we use
the Einstein relationship between mobility and diffusion
constant. The excess carrier concentration $n(x)$ in the
$p$ layer can then be found by solving

{\large
\begin{equation}
\label{transport}
\begin{array}{l l l l}
&\frac{d^{2}n}{dx^2} 
& +\frac{qE(x)}{k_{B}T}\frac{dn}{dx} - \frac{n}{L_n(x)^{2}}
& +  \frac{G(x,\lambda)}{D_{n}(x)}  \\
& =0 & \\
\end{array}
\end{equation}
} 

$L_{n}$ and $D_{n}$ are the electron
diffusion length and diffusion constant respectively.
$E(x)$ is the depth dependent effective electric
field due to the bandgap gradient. The expressions
for the band-gap are due to \cite{caseypanish78}
(direct gap) and \cite{hutchby76} (indirect).
The smaller contributions to the effective field from the bulk
photovoltaic effect, the mobility gradient and the Dember
potential are neglected.

The boundary condition at the surface is determined
by matching the drift and diffusion currents to the
surface recombination current. For the $p$ layer, this
takes the form

\begin{equation}
\label{pSurfaceBC}
\begin{array}{l}
\frac{qDE(x)}{KT}n(x) + qD_{n}(x)\nabla n(x)=S_{n}n(x) \\
$at $ x=0 \\
\end{array}
\end{equation}

where $S_{n}$ is the minority electron surface recombination
velocity. The second boundary condition in the depletion approximation
is that of zero excess minority carrier concentration at the edge
of the depletion region for $p$ respectively:

\begin{equation}
\label{piBC}
 n(x_{wp})=0   \\
\end{equation}

Equation \ref{transport} has an analytical solution
for ungraded samples with constant transport parameters
in the doped layers. For graded samples with
depth dependent transport characteristics, a standard 
numerical method is used.
The photocurrent $J_{p}(\lambda)$ from the p is given
by the diffusion current at the depletion edge

\begin{equation}
\label{pcurrent}
\begin{array}{l l}
J_{p}(\lambda)=qD_{n}\nabla n(x_{wp}) & \\
\end{array}
\end{equation}

where $x_{wp}$ is the position of the $p$
depletion edge. This expression assumes zero excess minority
carrier concentration at the edge of the junction within
the depletion approximation.

The validity of the this approximation on the minority
carrier gradient at the edge of the junction
was verified by analytically calculating the photocurrent
in the light at $x_{wp}$ for an ungraded sample with a high
current density. This was compared with the numerical result
for the same device. The largest error was was of the order
of 0.1\%  for a GaAs cell in the resulting photocurrent
from the $p$ layer.

Assuming 100\% collection efficiency, the current $J_{i}(\lambda)$
from the $i$ region is calculated from the integral of the generation
rate over the depleted regions

\begin{equation}
\label{iCurrent}
J_{i}(\lambda = q \int_{x_{p}-x_{wp}}^{x_{p}+X_{i}+x_{wn}}
\left[ G(x,\lambda) \right] dx
\end{equation}

The short circuit
current is then the sum of the contributions from the three regions

\begin{equation}
\label{TotalCurrent}
J_{sc}(\lambda=J_{p}(\lambda+J_{i}(\lambda+J_{n}(\lambda
\end{equation}

where the $n$ region photocurrent $J_{n}$ is calculated in a
similar fashion to $J_{p}$. The QE is defined in
terms of \jsc\ and the incident flux

\begin{equation}
\label{QE}
QE(\lambda)=\left[ \frac{J_{sc}(\lambda}{qF(\lambda)} \right]
\end{equation}

\subsection{Mirrors}
\label{mirrors}

The normal incidence mirror model treats the cell as a cavity
with uniform light intensity for a given wavelength. The
refractive index is an average over the structure.

The light intensity in the QWSC is calculated as a function of
wavelength by summing electric field amplitudes due to successive
reflections. The wavelength dependence of the front and back surface
reflectivities is neglected. A wavelength independent back surface
reflection phase change is included. The mean light intensity in
the cell is given by the squared modulus of the total electric
field amlpitude.
The QE enhancement above the well is neglected because
of the low levels of light reaching the back mirror at
these wavelengths.

\section{Modelling Method}
\label{modelling}

\subsection{QE model Parameters}
\label{parameters}

The main model parameters are the reflectivity, the absorption
coefficient,  $S_{n}$, $D_{n}$ and $L_{n}$.
Reflectivities were measured on a separate set of large area calibration
samples described in section \ref{samples}. An average
reflectivity is used in the modelling since the measured
data vary by no more than a few percent. Modelling of the absorption 
coefficient is described in
\cite{paxman93}.              

The surface recombination velocity $S_{n}$ is very dependent on 
sample growth and processing and is essentially used as a fitting
parameter at wavelengths below 400nm.

$L_{n}$ is also sensitive to growth and processing.
A wide range of values exist in the literature (\cite{ahrenkiel92},
cite{hamaker85}) and reliable published data can
only be found for $X$ compositions below approximately 40\%.
A number of simpler ungraded QWSC structures were grown at different
values of $X$ in order to increase our knowledge of $L_{n}$.
The following method was used to extract values of this parameter.

Inspection of the analytical solution to equation \ref{transport}
and equation \ref{pcurrent} shows that the expression for the
QE is independent of $D_{n}$ if $D_{n}$ is a constant. The only
free parameters in this case are $S_{n}$ and $L_{n}$. $S_{n}$
mainly influences the QE at short wavelengths whereas $L_{n}$
affects longer wavelengths. The ungraded $p$ layer QE can
therefore be modelled in terms of $S_{n}$ at short
wavelengths and $L_{n}$ at long wavelengths.

Fitting QE measurements of \pin\ and QWSC samples without grades
have enabled us to extract values of the diffusion length for
$X$ ranging from 20\% to 47\%. Since we have no samples outside
this range, we use the paramterisation in \cite{hamaker85}
for $X$ compositions above 47\%.

We use the values of $L_{n}$ descibed above to
model graded QWSC samples. This assumes that the $X$ dependance
of the diffusion length in the graded $p$ is similar to the
behaviour observed in separate ungraded structures with
different $X$ compositions.
In graded samples, however, $D_{n}$ is no longer a constant,
and has a significant effect on the QE.

Inspection of equations \ref{transport} and \ref{QE} shows that
for a graded sample the QE depends on the
gradient of the diffusion constant with $X$ composition
$\nabla_{X} D_{n}$ but not on the it magnitude of $D_{n}$.
The graded $p$ layer QE can therefore be modelled in terms
of $\nabla_{X} D_{n}$. In the absence of detailed $D_{n}$
measurements, we assume a constant gradient $\nabla_{X} D_{n}$
between the two Al fractions for each $p$ layer and use this
constant as the main fitting parameter.

\subsection{Mirror Parameters}
\label{mirrparams}

Modelling the mirrors involves three parameters.
We find that for low front surface reflectivities
the back surface reflectivity mainly determines the
level of QE increase in the wells. The amplitude
of Fabry-Perot oscillations is set by the
front surface. The phase change upon reflection
from the back surface is a poorly understood
quantity, but partly determines the position of 
Fabry-Perot peaks.

\section{Samples}
\label{samples}

The MBE samples were grown on a V80H Vacuum Generators
machine. The growth temperature
was at 630$^{\circ}$C. Temperature monitoring was carried
out using an optical pyrometer backed up by a substrate
thermocouple and RHEED observation of oxygen desorbtion
from the surface at 590$^{\circ}$C. The flux ratioes
(As:Ga 10:1) were measured with a beam monitoring
ion gauge. 

A series of ungraded 30 well QWSCs and control \pin\ structures
were grown by MBE at nominal Al fractions of 20\%, 30\% and 40\%.
These samples have 0.03$\mu$m windows grown at $X$=67\%. The Al fraction
is subject to increasing uncertainty up to about $X$=50\%
because of Ga source flux fluctuations. The controls are
identical in every respect except that the well material is
replaced by \algaas\ with the barrier Al fraction. A
$\rm Al_{0.2}Ga_{0.8}As$ double heterostructure \pin\
was also grown.

A set of three graded QWSCs was grown on the same MBE machine.
The $X$ compositions were calculated from the photocurrent spectra.
The Al fraction in the graded $p$ regions varies from $X$=44\%
at the front surface to $X$=22\% at the $p$-$i$ interface in sample
U4033. Analoguous grades in samples U4034 and U4035 ranged
from 67\% to 34\% and 67\% to 47\% respectively.

MOVPE growth details are given in \cite{roberts94}.
Sample QT468a is a 30 well ungraded QWSC with a 80\%
0.02$\mu$m \algaas\ window.

The samples were processed to circular 1mm gold
ring contact photodiode devices with a circular
600$\mu$m optical window. The devices are mounted
on TO5 headers.

The anti-reflection (AR) coating consists of 75nm of SiN.
Large area pieces of wafer from each sample were AR coated
to allow reflectivity measurements to be carried out.

Coating the back of a device with a mirror is achieved
by etching the substrate down to the $n$ layer, which acts as an
etch-stop. A metallic mirror is then evaporated  directly
onto the back surface of the $n$ region.

\begin{table*}
\label{lntable}
\begin{center}
\begin{tabular}{| c c c c c |}
\hline
Sample & Type     & Modelled $p$ Al & Diffusion       
& $S_{n}$ \\
       &          & fraction (\%)   & Length ($\mu m$) 
& $cm^{2}/s$ \\ \hline
U2027  & mqw      & 22              & 0.075            
& $10^{-5}$         \\ \hline
U2028$^{*}$  & pin DHet & 22              & 0.075            
& $10^{-5}$         \\ \hline
U4036  & pin      & 22              & 0.075            
& $10^{-5}$         \\ \hline
U2029  & mqw      & 35              & 0.06            
& $10^{-4}$         \\ \hline
U2030  & pin      & 35              & 0.05             
& $10^{-4}$         \\ \hline
U2031 & mqw      & 47              & 0.06             
& $5\times 10^{-4}$ \\ \hline
U2032  & pin      & 47              & 0.075            
& $5\times 10^{-4}$   \\ \hline
\end{tabular}
\end{center}
\caption[]{Modelled diffusion lengths for ungraded samples at three
aluminium fractions. The values of \protect $l_{n}$  show
remarkable consistency at low Al fractions. Poorer consistency
at high Al fractions is partly due to variable device performance.} 
\end{table*}

\subsection{Thin $p$ QWSC}
\label{thinp}

More detailed discussion of thin $p$ cells is given in \cite{paxman93}.
Preliminary studies indicate that series resistance has
a significant effect on unconcentrated AM1.5 performance
for thicknesses below about 0.1$\mu$m. 

\subsection{Determination of $L_{n}$ from \protect \\ Ungraded Devices}
\label{difflenmethod}

The model reproduces the QE of $pin$ and QWSC samples with very
similar values of $L_{n}$ and $S_{n}$.
These are given in table 1. Also shown in
the table are the band-gaps extracted from the photocurrent spectra.
We note that consistency between different types of samples is
very good for $X$=20\% but deteriorates at higher Al fractions.
Theory and QE data for the 30\% Al sample are given in
figure 2.

Modelling shows that $L_{n}$ 
decreases more slowly with increasing Al fraction than has
been reported in \cite{hamaker85}. It increases near
the direct - indirect transition in the region of 40\%
Al. This trend is consistent with published measurements
in the review article \cite{ahrenkiel92} although
our values are substantialy lower for reasons which are
not fully understood.

\begin{figure}
\label{ar2029}
\centerline{
\vbox{
\epsfxsize=2.6in
\epsffile{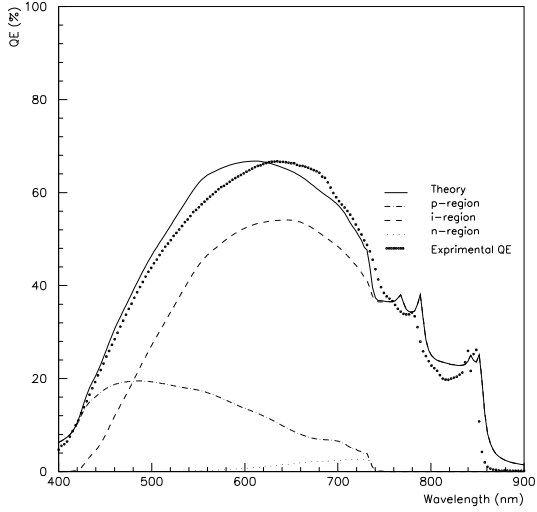}
\vspace{1ex}
}
}
\caption{Experimental data and model for a 33\% Al fraction 30 MQW solar
cell used to determine values of $L_{n}$.}
\end{figure}

\subsection{Determination of  $\nabla_{X} D_{n}$
\protect \\ from Graded QWSCs}
\label{dnmethod}

Graded samples were used to establish diffusion constant
gradients between the four different Al fractions and
are given in table 2.
The QE data and theory for the 33\% Al graded QWSC U4034
are given in figure 3.

\begin{figure}
\label{ar4034}
\centerline{
\vbox{
\epsfxsize=2.6in
\epsffile{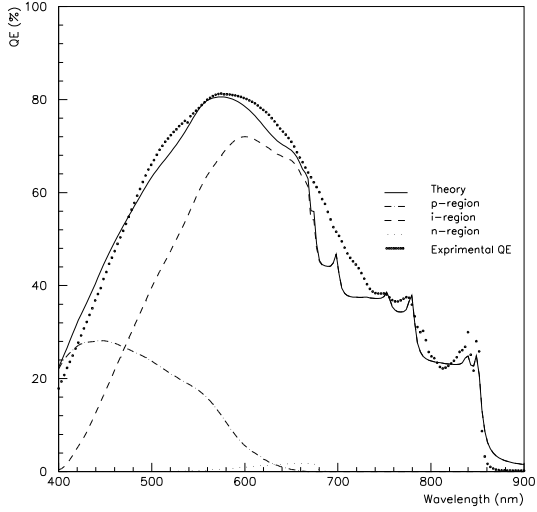}
\vspace{1ex}
}
}
\caption{Experimental data and model for a 33\% Al fraction 30
MQW solar cell incorporating a $p$ region compositionally graded
from 33\% to 67\% Al. Comparison with figure 2 shows
significantly improved QE at short wavelengths.}
\end{figure}

\begin{figure}
\label{m1fp468a}
\centerline{
\vbox{
\epsfxsize=2.6in
\epsffile{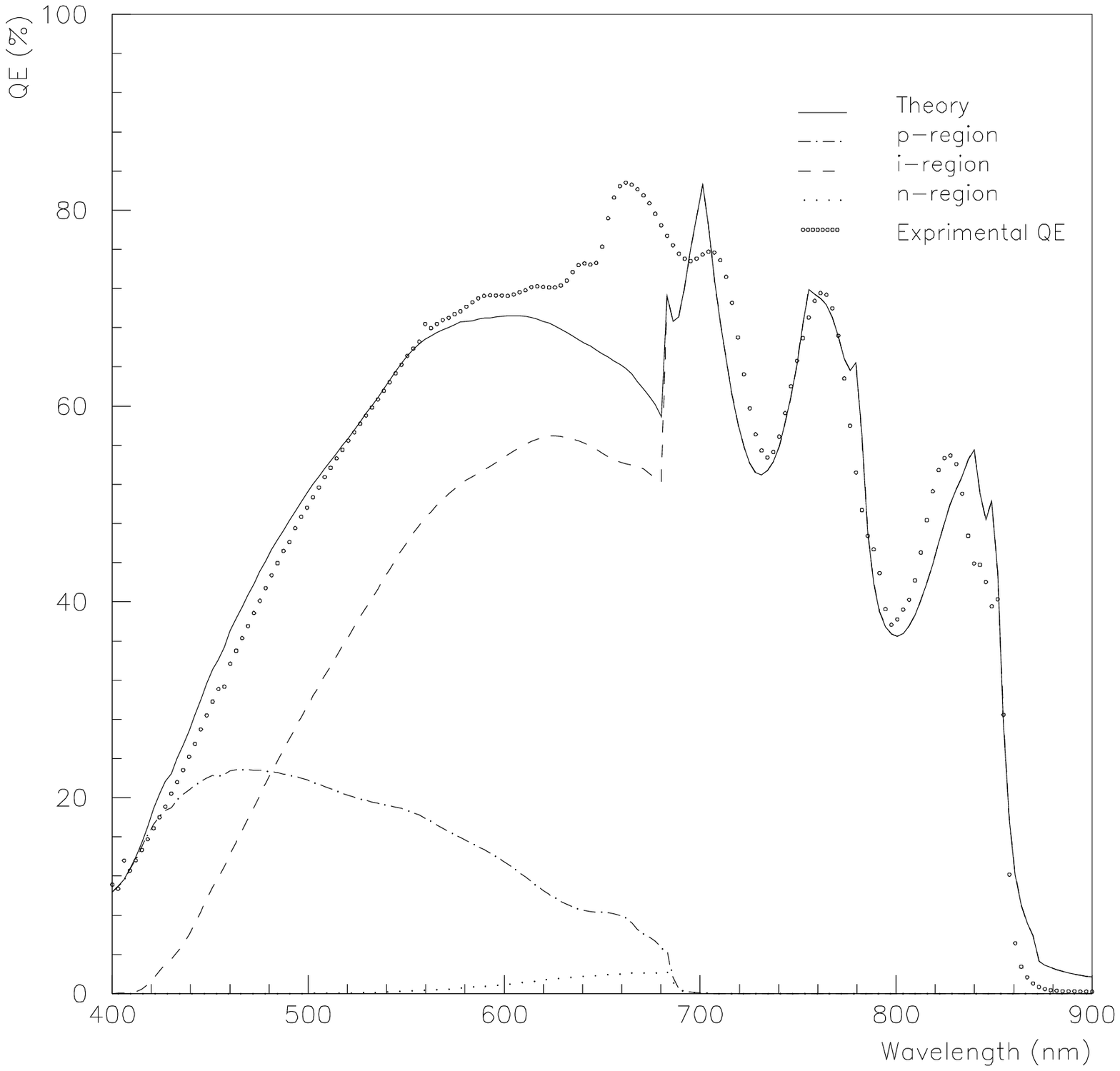}
\vspace{1ex}
}
}
\caption{mirrored 30\% cell QT468a showing Fabry-Perot effects. The short
circuit current enhancement for this device was 25\% overall.}
\end{figure}

Figure 5 shows the experiment and theory for the
mirror backed MOVPE sample QT468a. The integrated \jsc\ enhancement
for this device was  49\% in the well and of 28\% in \jsc\ overall.
Fabry-Perot peaks are visible, showing that front and back
surface interfaces are smooth. Other samples with accidentally
roughened back mirrors have shown higher \jsc\ enhancements. This
is attributed to non-specular reflection at the back surface which
increases the optical path length in the cell and hence improves
light absorption.

\begin{table*}
\label{dntable}
\begin{center}
\begin{tabular}{|c c c c|}
\hline
Sample & \multicolumn{2}{c}{$p$ grade}  & $\nabla_{X} D_{n} $ \\
       & Front Al fraction (\%) & Back Al fraction (\%) & (cm/s) \\
\hline
U4033 & 22 & 44 & $-5.5\times 10^{5}$ \\
U4034 & 34 & 67 & $-4.5\times 10^{5}$ \\
U4035 & 44 & 67 & $-4.3\times 10^{5}$ \\
\hline
\end{tabular}
\end{center}
\caption[]{Gradients of the diffusion constant in \protect \algaas
with respect to Al fraction. These values are derived from modelling
the QE of graded QWSC samples using the $L_{n}$ values given in table 1.}
\end{table*}
\section{Results}
\label{results}

\section{Optimisation}
\label{optimisation}

The model was used to design an optimised 50 well QWSC with a thinned
and graded $p$ layer. We chose to concentrate on a nominal barrier
aluminium fraction of 30\% and to base the design on our best
previous G951 cell which is described in \cite{paxman93}.

The p layer was thinned to 0.1$\mu$m. Modelling values of \jsc\
predict little current enhancement for grades with a top Al
fraction above about 44\%. We have therefore limited this optimisation to
44\% in view of increasing impurity incorporation at higher
Al fractions and greater uncertainty in modelling parameters
in this region.

The modelled \jsc\ under standard AM1.5 illumination
for a mirror backed device was 27.9 $\rm mA cm^{-2}$ for 
our contact design which has a 7\% shading loss.
Comparison with  G951 gives us an rough value for the efficiency
we expect from this sample. G951 has a \voc\ of 1.07V, a
fill factor of 78\% and a \jsc\ of 17.5$\rm mA\ cm^{-2}$. The
\voc\ is a little low for the optimised cell because of its
higher \jsc\ . If, however, we use these parameters to estimate
the efficiency of the optimised cell on the basis of the
modelled \jsc\, we obtain an efficiency of of 22.3\%.

This compares favourably with a GaAs cell described in
\cite{green94} which has an efficiency of 25.1\% for
a fill factor of 87\%, a \voc\ of 1.022V and a \jsc\
of 28.2$\rm mA\ cm^{-2}$. We note that the QWSC efficiency
would improve substantially if the fill factor QWSC were
increased.

\begin{figure}
\centerline{
\vbox{
\epsfxsize=2.6in
\epsffile{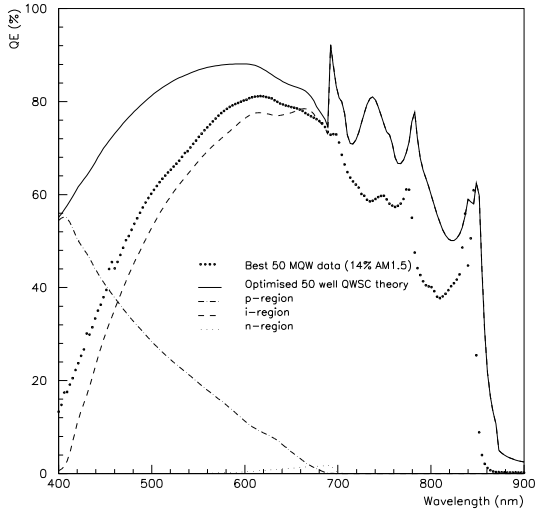}
\vspace{2ex}
}
}
\caption{Optimised 50 well QWSC consisting of a 0.1
$\mu$m $p$ region compositionally graded form 30\% - 44\% Al
and mirror backed. $I_{sc}$=25.5$mA (IEC-AM1.5)
cm_{-2}$. Also shown the experiment for the best  30\% QWSC prior
to optimisation.}
\end{figure}

\section{Conclusion}
\label{conclusion}

Previous work has demonstrated current enhancement and
promising voltage performance in QWSCs. Further theoretical
and experimental investigation has shown that useful
\jsc\ enhancements can be made in different wavelength
ranges. Improved design of the $p$ layer
can enhance the disappointing QE of \algaas\ cells
below 400nm, while a back mirror coating is seen to 
double the MQW current in 30MQW samples.

Theoretical predictions combining these improvements in
a single cell indicate that an \gaasalg\ cell can be
grown with efficiencies close to the unconcentrated
GaAs cells. Further improvements are expected if
fill factors in particular can be increased.
The design may be attractive for
the high bandgap component of a concentrator system,
either as an optimised \algaas\ cell, or a QWSC.
The QWSC may be attractive for this purpose
since its current output can be tuned to that of the lower
bandgap component by varying the number and/or width of
the quantum wells.


\end{document}